# ABSENCE OF "GHOST IMAGES" EXCLUDES LARGE VALUES OF THE COSMOLOGICAL CONSTANT

J. M. Eppley[1] and R. B. Partridge
Department of Astronomy, Haverford College, Haverford, PA 19041

ABSTRACT

We used the 1.4 GHz NRAO[2] NVSS survey to search for “ghost” images of radio sources, expected in cosmologies with a positive cosmological constant and positive space curvature. No statistically significant evidence for “ghost” images was found, placing constraints on the values of Λ, the space curvature or the duration of the radio-luminous phase of extragalactic radio sources.

## 1. INTRODUCTION

Cosmologists now accept the possibility of a non-zero value for the cosmological constant, Λ, which implies a small vacuum energy density in the Universe. A small positive value of Λ would help to resolve the "age problem" (some measures of the age of the Universe appear to exceed the time since the Big Bang); to explain the apparent acceleration of the expansion of the Universe (Perlmutter *et al.*, 1997; Reiss *et al.*, 1998; Garnavich *et al.*, 1998), or to ensure zero space curvature as favored by inflationary models even if the density of matter in the Universe is below its critical value, $\rho_c$. Values of Λ close to $2\ H_0^2 = (8\text{-}13) \times 10^{-36}\ \text{sec}^{-2}$ have been suggested (see Carroll *et al.*, 1992, for a review), corresponding to $\Omega_\Lambda = 0.67$ where $\Omega_\Lambda \equiv \rho_\Lambda / \rho_c$.

If Λ is somewhat larger, however, so that $\Omega + \Omega_m > 1$, the space curvature of the Universe becomes positive, even for very low matter densities, $\Omega_m$. We investigate one consequence of values of Λ close to or exceeding $3\ H_0^2$: light from a source can reach an observer from two opposite directions in a spatially closed Universe, as sketched in Fig. 1.

---

[1] Current address: Art Technology Group, 101 Huntington Ave., Boston, MA 02199.
[2] The National Radio Astronomy Observatory is a facility of the National Science Foundation operated under cooperative agreement by Associated Universities, Inc.

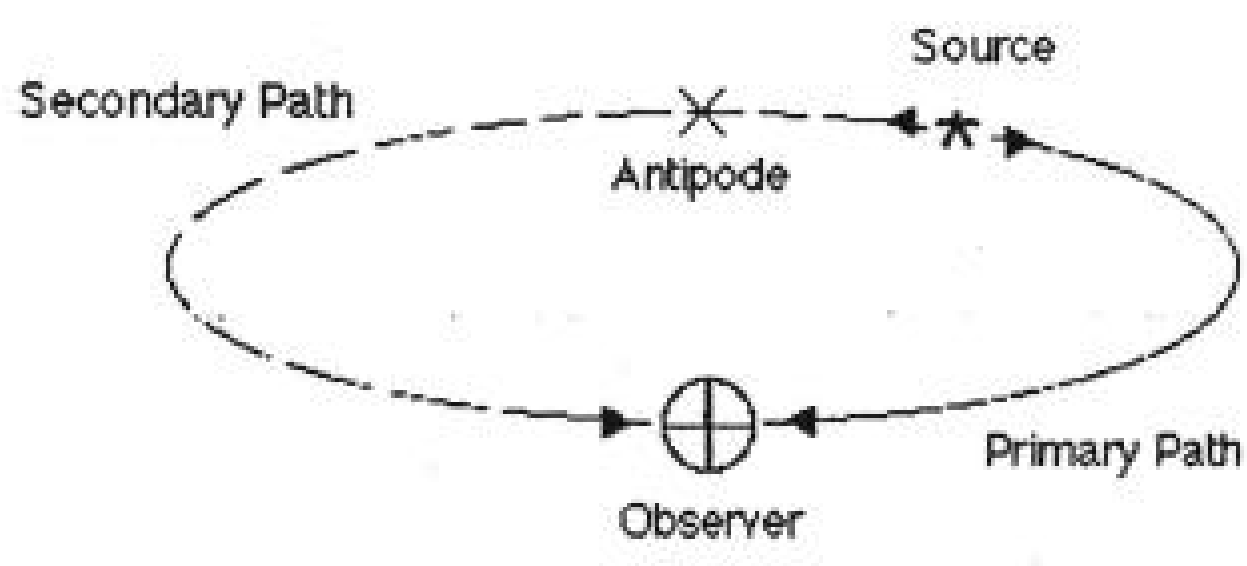

**Fig. 1 Formation of secondary or "ghost" images in positively curved space.**

The radiation leaving a (presumably isotropically radiating) source can take either the shorter path or the longer path around the curved geometry to reach the observer. The latter image is referred to as a “ghost image.” For closed, positively curved ($k = +1$) models with zero $\Lambda$, however, it may be shown (e.g., Petrosian and Salpeter, 1968) that the light travel time along the longer path exceeds the present age of the Universe, so no ghost image is formed. For closed models with $\Lambda > 0$, however, the age of the Universe is increased, allowing time in principle for light to travel along both paths to the observer. This is true also for the low-density, closed geometry model of Kamionkowski and Toumbas (1996). It is important to recognize that the "ghost image" of a particular source will not lie precisely at the opposite point of the celestial sphere, because light arriving to form the primary image and light arriving to form the "ghost image" will be emitted from the source at different times. Consequently, transverse motion of the source must be considered, in addition to relativistic aberration introduced by the motion of the observer.

One particular position of a source allows equal brightnesses for the primary and "ghost" images seen by a particular observer: the observer’s *antipode*. The distance from the observer to the antipode depends on the matter density $\Omega_m$ and $\Lambda$. Here, we choose to specify that distance in terms of the cosmological redshift $z_p$ of sources located at the antipode. For our test, sources need to be near the antipode, and the interesting range of $z_p$ is roughly 1-5, given the sources we use (c.f. Gott *et al.*, 1989, who use a lensed QSO to argue $z_p \geq 3.27$). Since the matter density is now thought to be quite low, say $\Omega_m \leq 0.3$ (see, e.g., White and Scott, 1996), the range of $\Lambda$ we can constrain is ~3-6 $H_0^2$ sec$^{-2}$, corresponding to a vacuum energy density contribution $\Omega_\Lambda$ ~ 1-2. It is already known that values of $\Lambda \geq 3\ H_0^2$ lead to larger probabilities of gravitational lensing than observed (see Fukugita *et al.*, 1992; Kochanek, 1996; or Carroll *et al.*, 1992, for a review). As will be seen, our independent method leads to the same qualitative conclusion: values of $\Lambda$ large enough to close the Universe are not consistent with our findings.

In Section 2, we assess previous searches for ghost images, then describe our improved procedure; in Sections 3 and 4, we explore consequences of our lack of detection of ghost images.

## 2. SEARCHES FOR GHOST IMAGES

Sources with redshifts in the range z = 2-5 have been known for decades. Many are QSO, which may not be suitable targets for such a search because of beaming (they may not be isotropic radiators). Optical searches employing galaxies, presumably approximately isotropic radiators, are hampered by confusion; the surface density of optical images is too high. Perhaps for these reasons, previous searches for ghost images by Solheim (1968), Petrosian and Ekers (1969) and Biraud and Mavrides (1980) have used radio sources.

### 2a. *Improved Search*

We too employ radio sources, but our work improves on these earlier efforts in two significant ways. First, we have a deeper, more uniform and more extensive radio survey available, the $2 \times 10^6$ entry NRAO VLA Sky Survey (NVSS) of Condon *et al.* (1998). In addition, we included relativistic aberrations (up to 8′) in the arrival directions of the rays of both the primary and the ghost image due to the velocity of the solar system. For some regions of the sky, this effect was large enough to place the ghost image outside the primary power beam of the telescopes used in previous searches. In our work, the aberration corrections were made assuming the motion of the barycenter of the solar system derived from COBE observations: v = 370 km / sec in the direction $\ell$ = 264.31° ± 0.16°, b = +48.05° ± 0.09°, given in Galactic coordinates (Lineweaver *et al*., 1996).

2b. *Our Procedure*

The 1.4 GHz NVSS Survey covers the entire sky north of –40° declination with 45″ resolution and a minimum detectable flux of ~2.5 mJy. We divided the database into 24 one hour strips in Right Ascension, with dec = –40° to +40°. A short program extracted an NVSS source, corrected its apparent position for relativistic aberration, and then calculated the direction to the ghost image, also corrected for aberration. We then examined the NVSS catalog to see if another apparent source were present in a set of circular zones about the predicted position of the ghost image; we took circular areas of radius 25″, 50″…200″ about the predicted position for each source, and counted the number of NVSS sources seen in each. The surface density of sources in the NVSS is large (~50 / $deg^2$), so we expect a substantial number of purely accidental coincidences. To control for these chance coincidences, we examined sets of concentric circular areas centered at positions *displaced* from the calculated position for the ghost image by ±5′ and ±10′ in RA and dec. Twenty four such control areas were examined.

We then added up the counts for all 1.37 million NVSS sources in the declination range –40° to +40°. In the absence of ghost images, we expect the number of purely accidental coincidences to scale as the square of the radius of the circle examined, if the distribution of sources is random. If ghost images are present, we expect higher counts in those circles centered on the correct ghost image position, causing a departure from the $r^2$ dependence at small r. Fig. 2a and 2b shows no evidence for such an effect. Counts in the control areas matched the expected $r^2$ dependence well.

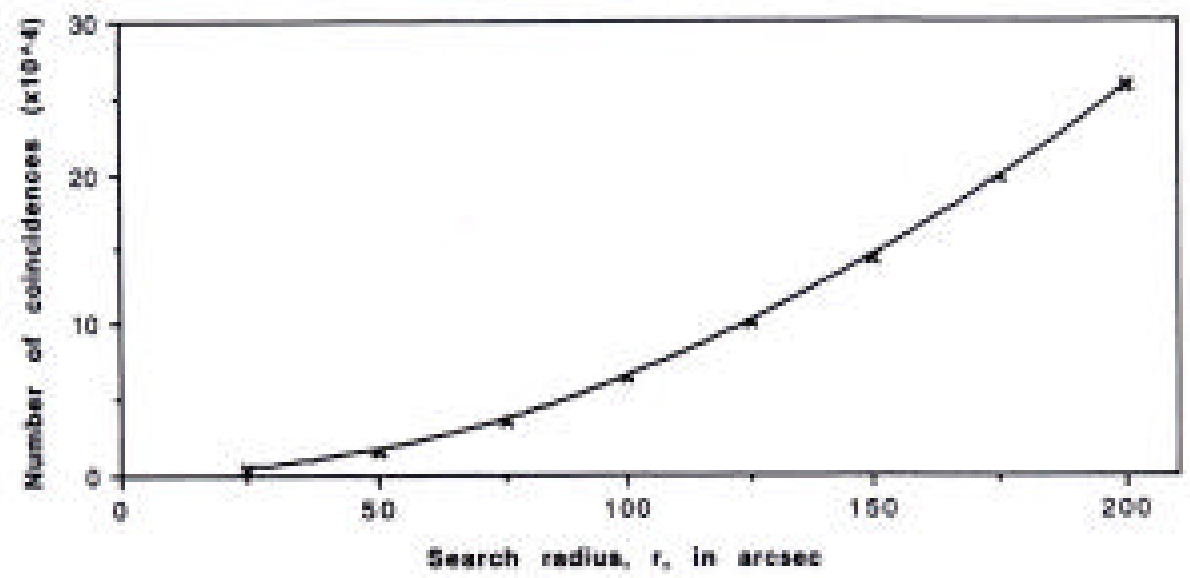

**Fig. 2a Upper panel: number of coincident sources detected in search areas of radius r. Note the essentially parabolic, $r^2$, dependence expected for a random distribution of sources.**

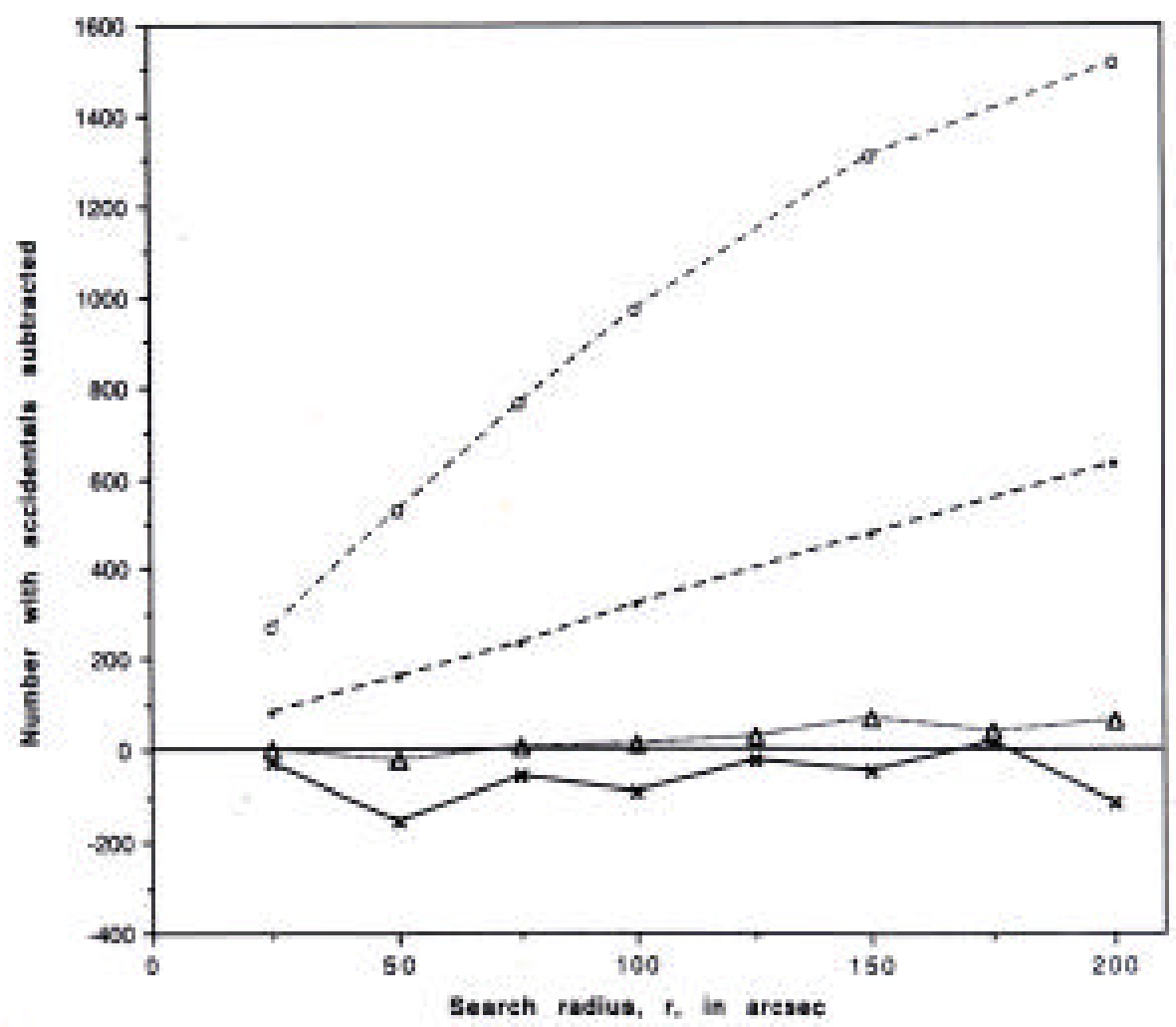

**Fig. 2b Lower panel, solid lines: number detected after accidental coincidences have been subtracted. The crosses are the counts centered on the expected position of ghost images; the triangles are counts in the twenty-four control areas. The dashed curves in the lower panel are our estimates of *one twentieth of* the counts of ghost images we would expect in a $k > 0, \Lambda > 0$ Universe. Clearly we see no such effect.**

2c. *Binning by Flux as a Proxy for Redshift*

Ghost images are expected only for sources relatively near the antipodal redshift. Few sources in the NVSS have measured redshifts, however, so we are unable to restrict our search to sources at a particular redshift. There is, however, some weak dependence of typical source flux with redshift. We thus used flux density as a crude proxy for redshift, and divided our sources into four flux density bins. No significant increase in the number of coincidences was found for any part of the survey examined at any flux range.

2d. *Restrictions on Flux Ratios*

We also performed a restricted search for ghost images by including only candidates with flux densities within a factor 3 (or 2) of the flux density of the primary image. We did so to reduce the number of accidental coincidences without eliminating many true ghost images. As Fig. 1 suggests, the travel time of the two rays is different unless the source happens to be located exactly at the antipodal redshift, $z_p$. Ghost images are expected only if the difference in travel time $\Delta t$ is less than the typical duration of the radio-active phase of the sources, which we take to be $\sim 10^8$ yr. Ghost images of radio sources will have distances, and hence flux densities, close to the primary images. Thus applying limits on the ratio of ghost flux to primary flux should help isolate true matches. When the restrictions in flux density were applied, fewer accidentals were detected, but even this restricted search yielded no significant evidence for ghost images.

## 3. IMPLICATIONS OF THESE NEGATIVE RESULTS

We will examine below the limits our negative results fix on $\Lambda$ and closed models. First, we need to take account of other astrophysical processes that might mask "ghost images" if they were in fact present.

Given the short lifetime of many radio sources (we assume $\Delta t \leq 10^8$ yr), only sources with redshifts near $z_p$ will produce detectable ghost images, as we saw above. This problem was recognized by Petrosian and Ekers (1969); following their work we estimate the fraction of NVSS sources we expect to possess ghost images. For $\Delta t \sim 10^8$ yrs and a flux density limit of 2.5 mJy, it is 2-3%, or ~30,000 sources in our work. Unlike some earlier work on ghost images, our search included a large enough number of sources to rule out even this low percentage of ghost images (at $>10\sigma$).

In addition, sources near the antipode are gravitationally lensed, and hence substantially magnified (Petrosian and Salpeter, 1968). We need to take account of the increase in the angular diameters of sources near the antipode; the magnification factor is several hundred, depending on $z_p$. Consequently, at the NVSS resolution of 45″, only sources with kiloparsec size or below will be detected; larger structures will be resolved out by the VLA. We thus could not detect classical, extended, radio galaxies, but could detect the numerous nuclear starbursting galaxies found in mJy surveys (e.g., Donnelley *et al.*, 1987).

Finally, there is the possibility that ghost images are indeed formed, but that the radio sources we used have substantial transverse velocities, $v_{tr}$, so that the direct and the "ghost" image are not diametrically opposed because the sources moved in the time interval $\Delta t$. As before, we take $\Delta t \sim 10^8$ yr to estimate the change in position, and assume that the transverse velocity of the radio sources can be represented by a Gaussian distribution proportional to exp $[1/2\ (v/v_0)^2]$. As a value for the dispersion $v_0$ in transverse velocities, we take

$$v_0 = \sqrt{\frac{2}{3}} \left( \frac{300 \text{ to } 500}{\bar{z} + 1} \right) \text{km/sec},$$

where the $\sqrt{2/3}$ arises because we consider only the transverse component, and (z + 1) enters because we assume the velocities are gravitationally induced and hence increases with time; $\bar{z}$ is the typical redshift of the sources. In Table 1, we evaluate

TABLE 1

FRACTION AND TOTAL EXPECTED COUNTS OF GHOST IMAGES WITHIN A CIRCLE OF RADIUS r OF THE EXPECTED POSITION

| radius, r | $v_0$ = 61 km/sec | | $v_0$ = 204 km/sec | |
|---|---|---|---|---|
| | fraction | number | fraction | number |
| 25″ | 0.16 | 5440 | 0.047 | 1600 |
| 50″ | 0.31 | 10540 | 0.095 | 3230 |
| 75″ | 0.45 | 15300 | 0.14 | 4760 |
| 100″ | 0.57 | 19400 | 0.19 | 6460 |
| 150″ | 0.77 | 26200 | 0.28 | 9520 |
| 200″ | 0.89 | 30300 | 0.37 | 12600 |

two illustrative and conservative cases, $v_0$ = 204 km/sec (with 500 km/sec and $\bar{z}$ = 1) and 61 km/sec (with 300 km/sec and $\bar{z}$ = 3). The change in angular position is then given by eq. (14) of Petrosian and Ekers (1969); in our notation,

$$\Delta q = \left(\frac{v_{tr}}{c}\right)\frac{3\Omega_m}{\left(\Omega_m + \Omega_\Lambda - 1\right)} \ .$$

With $\Omega_m \sim 0.3$ and $\Omega_\Lambda > 1$ as required for the formation of ghost images,

$$\Delta q \leq 3\left(\frac{v_{tr}}{c}\right) = 2\ \left(v_{tr}\right)\text{arcsec} \ ,$$

with $v_{tr}$ in km / sec. Now, for $v_0$ = 204 km / sec or 61 km / sec, we calculate the fraction of sources that have $\Delta\theta$ less than each of the radii we used in our search. For instance, with $v_0$ = 61 km / sec and a Gaussian distribution of transverse velocities, only 11% of sources would have $\Delta\theta > 200''$; that is, 89% of ghost images lie within a 200″ circle of the expected position. The radii of our search regions were selected to include a reasonable fraction of all ghost image candidates, even for $v_0$ as large as 204 km / sec, and at the same time to keep the number of accidental coincidences low. For a typical radio source lifetime of $10^8$ yrs, we expect 2-3% of all radio sources (that is, 34,000 in our work) to have detectable ghost images in a $\Lambda > 0$ closed geometry. This number, multiplied by the values in cols. 2 and 4 of Table 1 allows us to predict the signal we should see if ghost images are indeed present (cols. 3 and 5 of the table, and the dashed curves in Fig. 2 which shows *one twentieth* of the expected number of counts if ghost images are present). Actual counts of ghost images fall well below the predictions, and are consistent with zero.

## 4. LIMITS ON Λ AND SPACE CURVATURE

We turn next to limits on $\Lambda$, given our failure to find ghost images. First, we reemphasize that this test is of value only in positively curved cosmological models. For $\Omega_m \sim 0.3$, we thus cannot test values of $\Omega_\Lambda < 0.7$ or $\Lambda$ below 2.1 $H_0^2$ for any value of $z_p$. Our negative results, however, do rule out values of $\Lambda$ lying between 3 $H_0^2$ and about 4.5 $H_0^2$ ($\Omega_\Lambda$ = 1.0 to 1.5) for $\Omega_m$ = 0-0.3, and values of $\Lambda \geq 6\ H_0^2$ for $\Omega_m$ as large as 0.5. The method employed here cannot test smaller values of $\Lambda$ because the value of $z_p$ becomes too large, say $z_p \geq 5$ (Refsdal *et al.*, 1967). Hence this test does not constrain values of $\Lambda$ favored by groups using supernovae to determine cosmological parameters (Perlmutter *et al.*, 1997;

Reiss *et al.*, 1998). To avoid our constraints on $\Lambda$ would require either very short average lifetimes of radio sources ($\Delta t < 10^7$ yr) or large transverse velocities ($v_0 \geq 700$ km / sec).

Our failure to observe ghost images also sets constraints on the closed, low-density cosmological model constructed by Kamionkowski and Toumbas (1996). Their model has a closed spatial geometry even when the density of non-relativistic matter $\Omega_m$ is less than 1. The model requires a scalar field with an energy density that varies with the inverse square of the scale factor and has a present equivalent density $\Omega_t$. In this model, the presence of the cosmic microwave background (CMB) is explained by placing the surface of last scattering at a polar-coordinate distance of $n\pi$, with n an integer. If n = 1, the last scattering surface is at the antipode, so $z_p \sim 1100$. Our results do not constrain this variant of the model. The more interesting case is n = 2, in which case the CMB photons travel all the way around the closed geometry, and "… when we observe the CMB, we are looking at the *local* (rather than distant) region of the Univese as it was at a redshift z ~ 1100" (Kamionkowski and Toumbas, 1996). For this case with n = 2, the value of $\Omega_t$ is determined by $\Omega_m$:

$$\Omega_t = \left[ \frac{2p\sqrt{1-\Omega_m}}{\operatorname{arcsinh}\left(2\sqrt{1-\Omega_m}/\Omega_m\right)} \right]^2 + 1 - \Omega_m$$

(eq. 5 of Kamionkowski and Toumbas, 1996). For $\Omega_m$ = 0.1, 0.3 and 0.9, for instance, $\Omega_t$ = 3.6, 5.4 and 9.3, respectively. The antipode, at a polar-coordinate distance $\pi$, is found at a redshift $z_p$ given by:

$$p = \sqrt{\Omega_m + \Omega_t - 1} \int_0^{z_p} \frac{dz}{\left[ \Omega_m (1+z)^3 + (1-\Omega_m)(1+z)^2 \right]^{1/2}}$$

(*op cit*, eq. 4). The absence of ghost images in the NVSS survey implies $z_p > 5$. That allows us to rule out these models for values of $\Omega_m \geq 0.23$. If n > 2, the constraints are even more stringent. Kamionkowski and Toumbas (1996) note that other astronomical evidence also casts doubt on $n \geq 2$ variants of their model, but do not provide a complete analysis.

Early work on this project at Haverford was carried out by Todd Green. Research supported in part by the US National Science Foundation and the Wm. Keck Foundation through a grant to the Keck Northeast Astronomy Consortium. We also thank an anonymous referee for a very helpful suggestion, and many colleagues for useful comments on the draft of this paper.

**Figure Captions**

1.) Formation of secondary or “ghost” images in positively curved space.

2.) Upper panel: number of coincident sources detected in search areas of radius r. Note the essentially parabolic, $r^2$, dependence expected for a random distribution of sources. Lower panel, solid lines: number detected after accidental coincidences have been subtracted. The crosses are the counts centered on the expected position of ghost images; the triangles are counts in the twenty-four control areas. The dashed curves in the lower panel are our estimates of *one twentieth of* the counts of ghost images we would expect in a $k > 0$, $\Lambda > 0$ Universe. Clearly we see no such effect.